\documentclass{aa}
\usepackage{graphics}

\begin{document}  
\thesaurus{04               
		(04.01.2;   
		 08.01.1;   
		 08.01.3;   
		 08.06.3)}  

\title{On-line determination of stellar atmospheric parameters $T_\mathrm{eff}$, $\log g$, [Fe/H]
from ELODIE echelle spectra. \\
II - The library of F5 to K7 stars.}
   
\thanks{based on observations made on the 193cm telescope at Observatoire
de Haute-Provence, France. The library of spectra and corresponding data are only  available in 
electronic form at the CDS via anonymous ftp to cdsarc.u-strasbg.fr (130.79.128.5), or via 
http://cdsweb.u-starsbg.fr/Abstract.html.}

\author{C. Soubiran \inst{1}, D. Katz \inst{2}, R. Cayrel \inst{2}}

\offprints{Caroline Soubiran,\\ 
soubiran@observ.u-bordeaux.fr }

\institute{ 
Observatoire de Bordeaux, BP 89, F-33270 Floirac, France
\and 
Observatoire de Paris, DASGAL, 61 avenue de l'Observatoire, F-75014 Paris, France}

\date{Received ; accepted }

\maketitle

\begin{abstract}
A library of 211 echelle spectra taken with ELODIE at the Observatoire de 
Haute-Provence is presented. It provides a
set of spectroscopic standards covering the full range of gravities and metallicities in 
the effective temperature interval [4000 K, 6300 K]. 
The spectra are straightened, wavelength calibrated, cleaned of cosmic 
ray hits, bad pixels and telluric lines. They
cover the spectral range
[440 nm, 680 nm] with an instrumental resolution of 42000. For each star, basic
data were compiled from the Hipparcos catalogue and the Hipparcos Input Catalogue. 
Radial 
velocities with a precision better than 100 $\mathrm{m.s}^{-1}$ are given. Atmospheric parameters 
($T_\mathrm{eff}$, $\log g$, [Fe/H]) from the literature are discussed. Because of scattered determinations
in the bibliography, even for the most well-known stars, 
these parameters were adjusted by an iterative process which takes account of 
common or different spectral features between the standards, using our homogeneous set of spectra. 
Revised values of effective temperature, gravity and metallicity are proposed. They are still consistent with
the literature, and also lead to the self-consistency of the library, in the sense that similar spectra 
have similar atmospheric parameters. This adjustment was performed by using step by step 
a method based on the least square comparison
of carefully prepared spectra, which was originally developed for the on-line estimation of 
the atmospheric parameters of faint field stars ( companion paper in the main journal).
  

\end{abstract}


\section{Introduction}
The library presented in this paper was compiled in order to provide a grid of reference spectra for
the determination of reliable atmospheric parameters from high-resolution spectra taken with the echelle 
spectrograph ELODIE at the Observatoire de Haute-Provence. It was the
first step of a chemical and kinematical probe of the Galaxy that we have been conducting for five
years with different spectrographs, and for which we have recorded 600 spectra 
of field stars at various resolutions and signal to noise ratios (S/N).
Some 132 of our brighter target stars ($11\leq V \leq 14$) have
been observed with ELODIE, and we have also observed in the same conditions, but higher S/N, 211 among the most
studied spectroscopic standards, including the Sun, in the temperature interval [4000 K, 6300 K] to build up a
library of reference spectra. 
 This temperature interval has been 
chosen because it covers the
full span of stars born from the beginning of the Galaxy. The metallicity of these
stars is supposed to  
reflect  the metallicity of the interstellar material from which they were formed, tracing galactic evolution. In hotter stars, the metallicity might be altered by physical
processes, and spectra of cooler stars are crowded by molecular bands and might lead to
erroneous results. Classical methods estimating [Fe/H], where equivalent widths of weak 
lines are measured, are not applicable on low S/N spectra. Furthermore the metallicity cannot be extracted 
independently of the
two other atmospheric parameters on which the spectrum depends : effective temperature $T_\mathrm{eff}$, and 
gravity $\log g$.
 Therefore, in order to estimate at the same time the three parameters,
we have developed a method and a software, TGMET, relying on the
least-square comparison of a target spectrum to the library of reference spectra for which the
atmospheric parameters are well-known. A version of the software is now available for 
observers working on ELODIE at the Observatoire de Haute-Provence, and allows on-line estimation
of ($T_\mathrm{eff}$, $\log g$, [Fe/H]) for stars in the spectral range F5 to K7, observed at a S/N ratio
greater than 10. The method is fully described in a 
companion paper (Katz et al. 1998, paper I). 
\\
 
              In order to use the library as a reference grid, one has
to know precisely  the atmospheric parameters of all the standards. This is not a simple task, 
since even for stars which
have been studied at high resolution and high S/N by several authors, the results of detailed 
spectral analyses 
can differ significantly due to the use of different
models of atmospheres and comparison stars. Section 3 is devoted to the problem of finding the
best atmospheric parameters for the standards, which must be consistent with the literature and
also with spectral analogies in the library. Reliable determinations of ($T_\mathrm{eff}$, $\log g$, [Fe/H]) were
compiled from the literature, but a simple average
of these values was not satisfactory, and for several stars, disagreements were found between the 
mean atmospheric parameters
and their spectrum. The avalaibility of an homegeneous set of high resolution, high S/N, wide
wavelength range spectra led to the obvious idea that it could help finding the most consistent parameters, in 
the sense that stars having
very similar spectra should have very similar atmospheric parameters. As TGMET is a tool 
which was conceived especially to measure similarities and discrepancies between spectra, it was used
iteratively  taking in turn each
reference star as an unknown object to adjust its parameters with respect to the rest of the library. 
Finally an 
homogeneous set of parameters and a self consistent library were obtained, as described in Sect. 3. \\

This library can be used for several astrophysical purposes like calibrations, 
differential studies of lines,
or spectral classification. As TGMET includes  very careful preparation of the
spectra ( continuum flattening, spurious radiation events removal, etc...), the spectra of the library  
are ready to be used. 
 The observational material and preparation of the spectra are briefly described 
in Sect. 2, as well as the basic data of these stars :
coordinates, magnitudes (including the bolometric absolute magnitudes), spectral types, distances, 
radial velocities, spatial velocities.     

\section{Observational material, TGMET, basic data}

The reader is encouraged to read paper I to gain a detailed description of ELODIE spectra
and TGMET.
The TGMET software has been installed at the Observatoire de Haute-Provence to provide 
astronomers observing with ELODIE an on-line estimation of the atmospheric parameters of their 
stars if
they fall in the correct temperature interval.
At a mean S/N of 100, an internal accuracy of 85 K, 0.28,
 0.16  is obtained on $T_\mathrm{eff}$, $\log g$ and [Fe/H] respectively. 
 TGMET runs in two phases : preparation of the target spectrum and comparison to the library
of reference spectra. All the reference spectra underwent the first phase, the main lines of which are
recalled here. The second phase of TGMET was performed on the standards in a way which is described in Sect. 3. 
The ELODIE spectrograph and the estimation of radial 
velocities by cross-correlation are presented in Baranne et al (1996) and in the {\it Inter-Tacos}
user's guide (Queloz 1996). For the kind of stars we are dealing with, the radial velocities have
an accuracy better than 100 m.s$^{-1}$. They are given in Table 1 (only available at the CDS). 
ELODIE spectra cover the interval [390 nm, 680 nm] over 67 orders. Because of the under-illumination of
the bluest orders, we only kept 47 orders covering the range [440 nm, 680 nm]. 
Each extracted spectrum was treated 
to remove all the features which are not intrinsic to the star. This treatment includes the
correction of the blaze efficiency, thanks to a
polynomial fitted on the spectra of several very metal-poor 
stars, and the removal of cosmic ray hits and telluric 
lines.\\

The 211 standards were observed between April 1994 and January 1998 with a S/N at
550 nm ranging between 36 and 381. The average value of S/N is 120, but lower S/N 
correspond to faint metal-poor stars kept in the library to have a good coverage of
the full metallicity range. Some spectra have also been kindly made available
by several observers.\\

We compiled basic data which can be
helpful for people who want to use the library. All the stars of the library, except the Sun,
 belong to the
Hipparcos catalogue from which coordinates, proper motions and parallaxes
were taken. Spectral types and visual magnitudes are from the  Hipparcos Input Catalogue. 
For 192 stars which
have a relative error on parallax lower than 30\%, absolute magnitudes, distances 
and one sigma error bars were computed. The bolometric correction was applied according to the relation
depending on Teff and [Fe/H] established by Alonso et al. (1995). For 5 stars too cool for this
correction, the M$_v$ absolute magnitude is given instead of Mbol.
The 3 components (U,V,W) of spatial velocities with respect to the Sun were also computed.
Basic data are listed in Table 1, only available at the CDS, together with the atmospheric
parameters described in the next section.\\

 To enable astronomers to make use of this library, the 211 spectra 
are available in FITS format at the CDS via anonymous ftp or WWW. Each
spectrum is made of a file of $1024 \times
47$ wavelengths with their corresponding flux (not normalised) plus a header containing observational information. The pixels
eliminated through the straightening and cleaning processes are flagged by a value of the corresponding flux of -100. No attempt was made to recombine the 47 orders in a single spectrum.

\section{ Atmospheric parameters}

\subsection{ Input parameters}
The vast majority of the reference stars which form the library were selected 
in the catalogue of
[Fe/H] determinations (Cayrel de Strobel et al. 1997).  A few others are from the list of
proper motion stars of Carney et al. (1994). The [Fe/H] catalogue provides near 6000 determinations
of ($T_\mathrm{eff}$, log g, [Fe/H]) from detailed analyses of high resolution,
high S/N spectra for 3248 stars. For each star, several different values are listed for their atmospheric
parameters. The difficulty of finding the "true" parameters for a given star can be illustrated by the 
case of the well-known deficient sub-giant HD 140283. Its first detailed analysis was performed by 
Chamberlain and Aller (1951), and since
then 28 detailed analyses have been reported, quoting $T_\mathrm{eff}$ from 5362 K to
6300 K, $\log g$ from 3.2 to 4.8 and [Fe/H] from -1.04 to -3.06, with standard deviations of 111 K, 0.39 and
0.23. A simple average is not correct since all the
analyses do not have the same weight. It is also worth noting that $T_\mathrm{eff}$ and $\log g$ quoted in
detailed analyses are often the values of the model atmosphere chosen to deduce the iron abundance and
are not obtained directly from spectroscopy. It is thus quite difficult
to know which parameters should be adopted for a given star. The [Fe/H] catalogue is complete up to
december 1995, and several recent references including new atmospheric parameters have been added 
to our list of determinations: Carney et al. (1994),  
Alonso et al. (1996b), Pilachowski (1996), Gratton et al. (1997), Nissen \& Schuster (1997), Nissen et al. (1997), 
Th\'evenin (1998). The sample of Carney et al. is not in the [Fe/H] catalogue because the 
metallicity estimations rely on low S/N spectra. The sample of Nissen et al. includes photometric metallicities and surface gravities derived from Hipparcos parallaxes. The study of Alonso et al. is not based on spectroscopy but 
concerns only effective temperatures
calibrated with the InfraRed Flux Method. As temperature is the parameter which
shows the largest scatter between the authors, two independant calibrations
of $T_\mathrm{eff}$ were also used : $T_\mathrm{eff}$ versus V-K and [Fe/H] and $T_\mathrm{eff}$ versus b-y, c1,
[Fe/H] (Alonso et al. 1996a). The V-K colour indices were found for 70 stars in the catalogue of 
Morel \& Magnenat (1978). Both b-y and c1 were found
for 143 stars in the catalogue of Hauck \& Mermilliod (1998). The references which appear the 
most often 
for this sample are Th\'evenin (1998) for 128 stars, Gratton et al. (1997) for 70 stars, McWilliam (1990) for 62 stars,
Alonso et al. (1996b) for 64 stars, Axer et al. (1994) for 32 stars, Pilachowski (1993) for 27 stars, 
Edvardsson et al. (1993) for 24 stars, Tomkin et al. (1992) and Luck \& Challener (1995) for 19 stars.  
The determinations of these different authors were compared to uncover any systematic offsets in
the various photometric and spectroscopic data sets. 
 There were none: values for individual stars present scatter , but there are no systematic trends 
between these authors. \\


\begin{figure}
\resizebox{8cm}{!}{\includegraphics{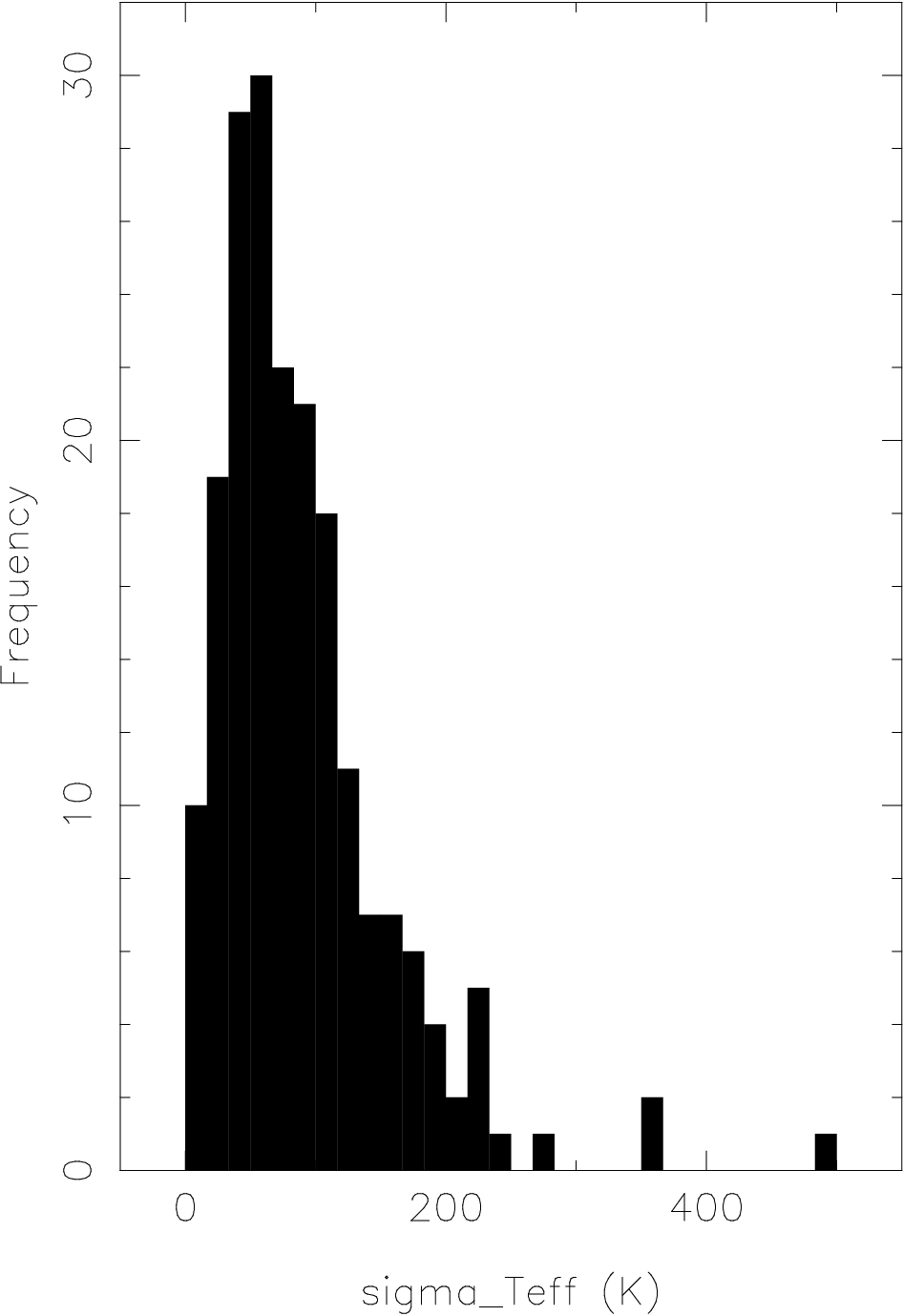}}
\caption{Histogram of the rms of the determinations of $T_\mathrm{eff}$ from the literature 
and photometry for the 211 standards.}
\label{}
\end{figure}

We first computed a weighted mean of the different determinations with higher weight for
recent determinations (after 1990) and lower weight for old ones (before 1980 when solid 
state detectors were not yet available). The weighted 
root-mean square (rms) of the different determinations measures their agreement. The histogram of 
the rms
for each parameter is shown respectively on Fig. 1 to 3. Figure 4 shows the histogram of the number
of [Fe/H] determinations per star. For the majority of
the standards the determinations are in a reasonable agreement, but large discrepancies can also be seen
for several stars. The situation is particularly worrisome for temperatures which appear to be that 
parameter the most
delicate to determine. As it is difficult to find an objective criterion to keep some determinations
and to eliminate others, we had to think of another way to find the best parameters for the
reference stars. 
The availability of high S/N spectra at the same resolution for all these stars provided a useful
support to perform an internal   
check of the library by comparing standards of the same type, and to adjust their parameters 
until reaching the self-consistency of the 
library. The method TGMET developed for the on-line determination of atmospheric parameters of 
anonymous stars, presented in
paper I, offered the opportunity to realise the comparison between the spectra in a quantitative way.


\begin{figure}
\resizebox{8cm}{!}{\includegraphics{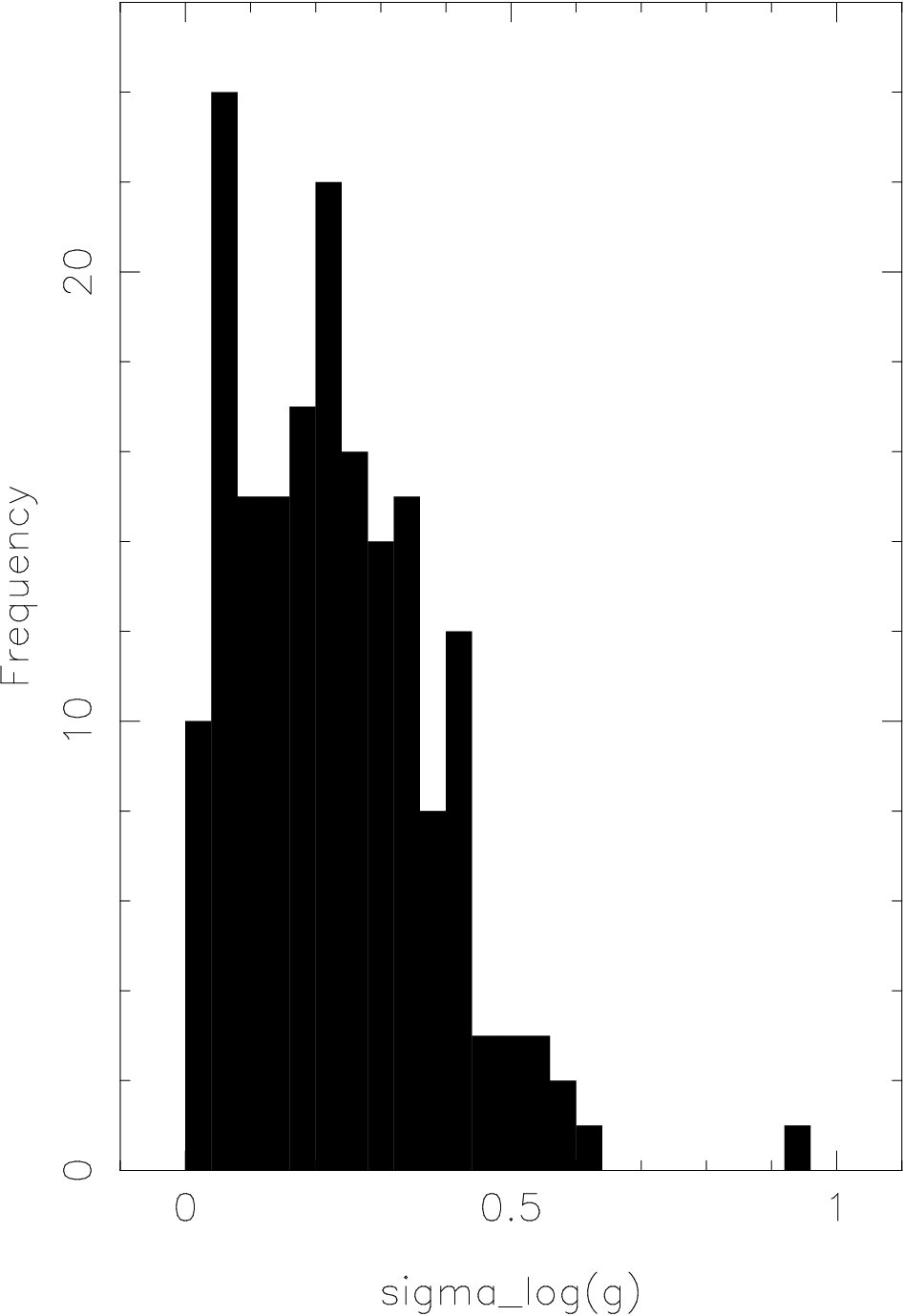}}
\caption{Histogram of the rms of the determinations of $\log g$ from the literature 
for the 211 standards.}
\label{}
\end{figure}


\begin{figure}
\resizebox{8cm}{!}{\includegraphics{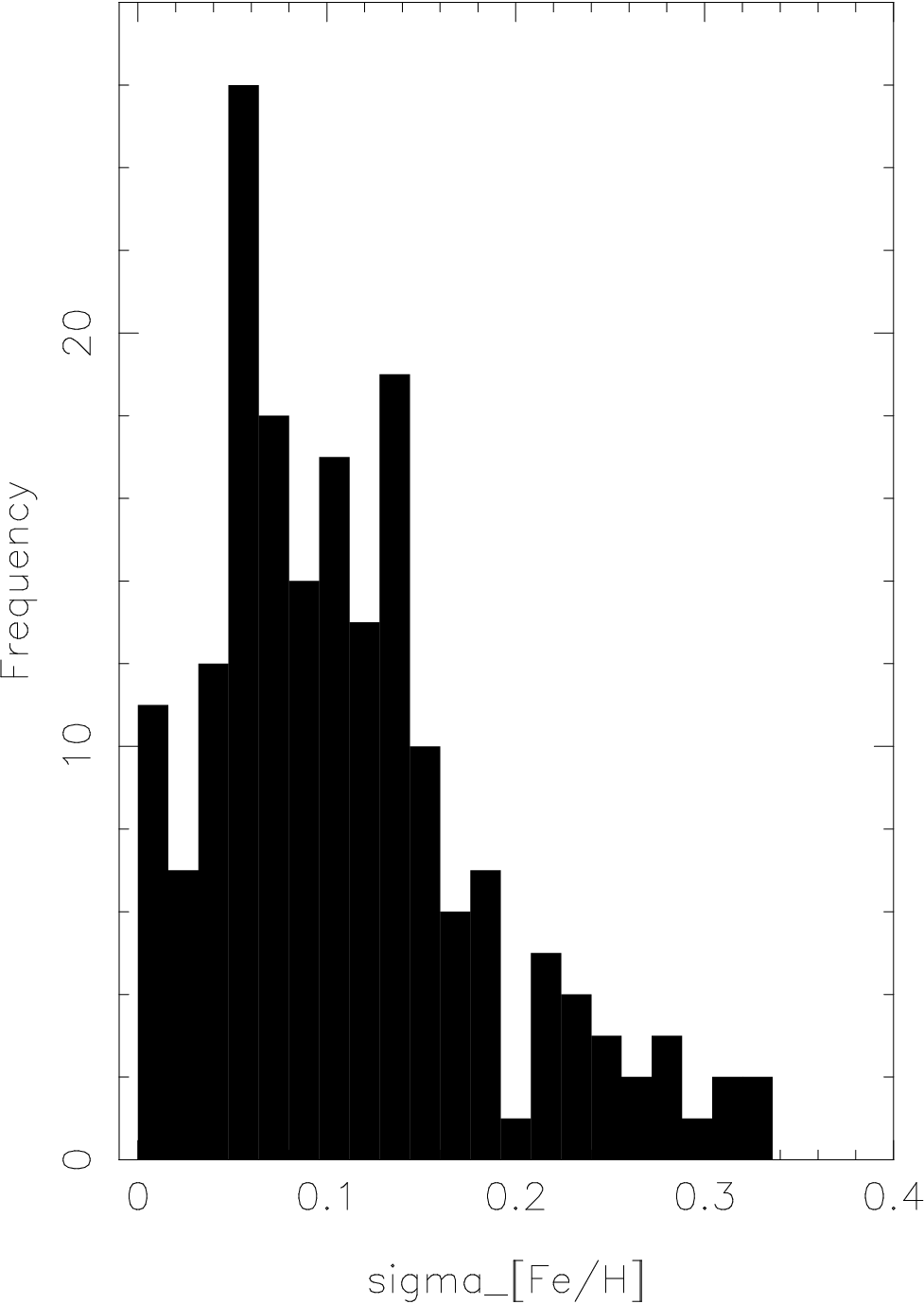}}
\caption{ Histogram of the rms of the determinations of [Fe/H] from the literature 
for the 211 standards.}
\label{}
\end{figure}


\begin{figure}
\resizebox{8cm}{!}{\includegraphics{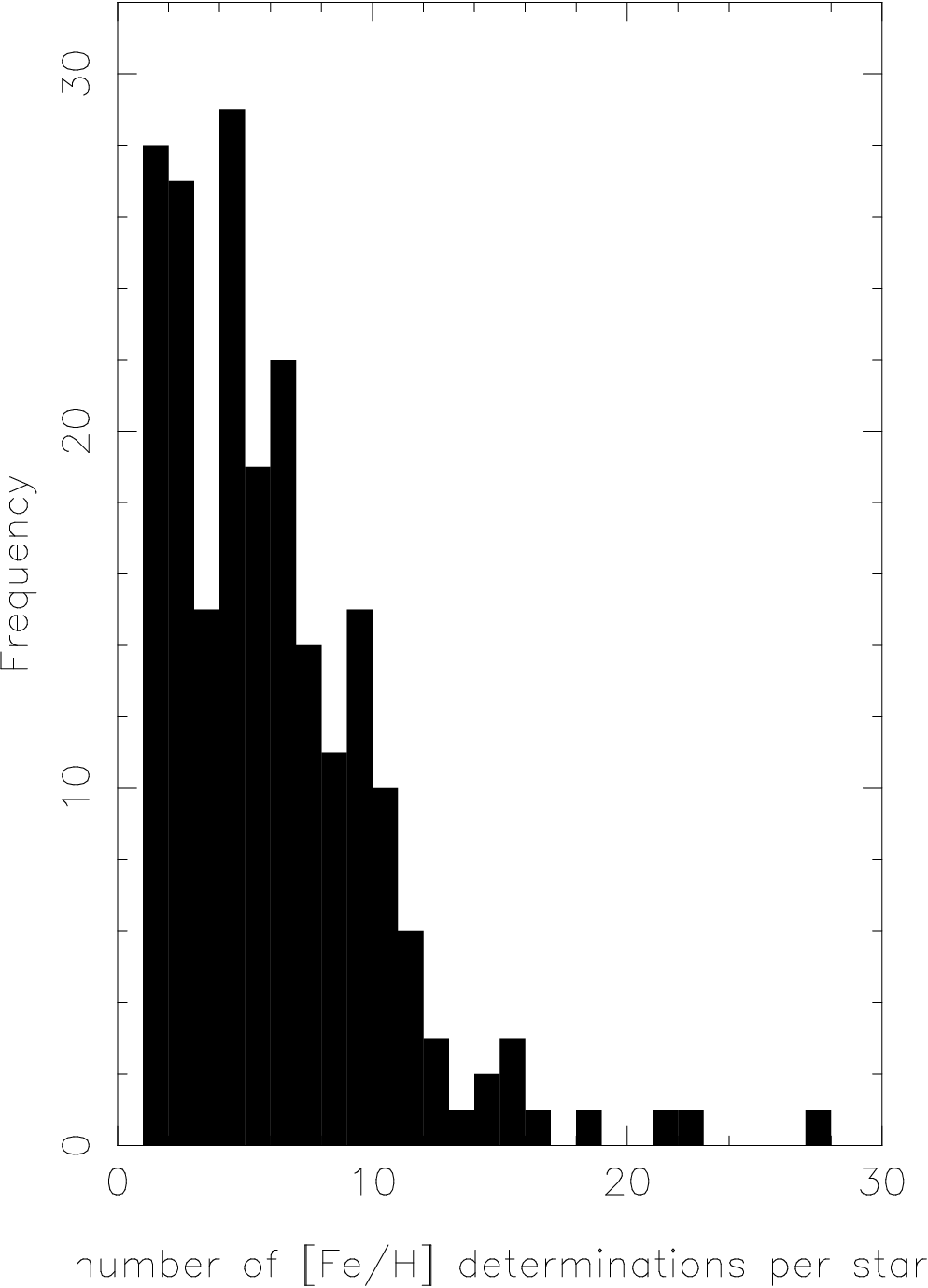}}
\caption{Histogram of the number of [Fe/H] determinations per standard.}
\label{}
\end{figure}

\subsection{ Final parameters}

The TGMET software, presented in paper I, provides a criterion of resemblance between two
spectra and an estimation of the atmospheric parameters of a target star. 
The estimation is performed by comparing the target spectrum
to each reference spectrum following
three steps. 
The first one is the convolution of all  spectra to exactly the same
resolution. 
 The degradation of the resolution is performed to 
eliminate the effects of
different projected rotational velocities when comparing two stars, and slight modifications of instrumental 
resolution
between observational runs. A value of 13 $\mathrm{km.s}^{-1}$ (FWHM) was chosen because it corresponds to the reference star having the lowest resolution (instrumental + intrinsic).
The wavelengths of the reference spectrum are then shifted to make its absorption 
lines coincide with those of the target spectrum.
The third step is a flux adjustment, order by order (on the 15 most significant ones), pixel by pixel, by the 
least square method. We use a criterion of resemblance, which is equivalent in practice to the reduced $\chi^2$.
The reference stars presenting  the lowest reduced $\chi^2$ are kept for the solution. The estimated parameters of the
target star are given by the weighted mean of the parameters of the best reference stars. 
The aim of the method is eventually to find in the library 
a twin of the target star, or several standards with very similar spectra. \\

When testing each
reference spectrum against the rest of the library, it was found in a few cases 
that spectra with significantly
different atmospheric parameters were found to be very similar. As the atmospheric parameters of the
reference stars are the result of the weighted mean described in the previous paragraph, it is
not surprising to have such discrepancies. Standard deviations of 145 K in
$T_\mathrm{eff}$, 0.35  in $\log g$ and 0.18  in [Fe/H] were obtained for the
distribution of the "mean literature" parameters versus "TGMET" parameters, with extreme discrepancies 
reaching 421 K, 1.27  and 0.49 . The discrepancies have two main sources.
The first one is the inhomogenity of the parameters from the literature, the mean of which can
lead to erroneous values, as described in the previous section. The second one is the inhomogenity of
the distribution of the reference stars in the 3D space of the parameters. The library is not a 
perfect 3D grid with equal spacing between the points. There are several sparse parts, especially among
metal-poor
stars and cool dwarfs. These holes in the library correspond to the absence of certain kinds
of stars in the [Fe/H] catalogue which are difficult to observe or to analyse. When it occurs in these parts of the library, a given star might
not have an analogue, and its nearest "neighbour" might be quite far. 
If this source of discrepancy can only be eliminated by filling the library with new reference stars of various parameters, 
the problem of inhomogeneous parameters can be partially solved by using  TGMET iteratively on each standard. \\
By slightly modifying the parameters of the reference stars, step by step,  
identical 
parameters should be found for twin spectra. The correction which can be applied to the initial parameters 
depends
on the consistency of the determinations in the literature and on the proximity of the neighbours. 
No correction was applied in three cases. The first one concerns the only star which is supposed to have perfectly known parameters : the Sun. 
The second case concerns 22 stars which do not have close neighbours in
the library, or whose parameters put them at the edges of the 3D space.  
The third case concerns stars for which the solution given by TGMET was satisfactory,
 that is the 
difference between the solution and the library parameters 
was smaller than 100 K, 0.5  and 0.3  respectively for $T_\mathrm{eff}$, $\log g$ and [Fe/H]. 
These values were adopted as a reasonable scatter inherent to the method and the library.
A limit to the correction was also fixed, depending on the reliability of the initial 
parameters.  
A level of quality was attributed to the starting point : high for many recent determinations in a good agreement,
medium for several determinations with a significant scatter or very few recent ones,
poor for old determinations
or large scatter.  There are only 8 standards of high quality, with at least
6 recent determinations showing a dispersion lower than 75 K, 0.4 and 0.15 in $T_\mathrm{eff}$, 
$\log g$ and [Fe/H].The corresponding limit of correction in $T_\mathrm{eff}$ was
100 K for good reference stars, 200 K for medium ones and 250 K for poor ones. In all cases
the maximal correction in $\log g$ was 1. and 0.5 in [Fe/H] with respect to the initial values. 
The corrections were performed step by step
by computing the difference between the input parameters and the output parameters. The steps 
were respectively
50 K, 0.1 and 0.05 for $T_\mathrm{eff}$, $\log g$ and [Fe/H].
The parameters converged after 6 iterations. Among the high quality standards, only 3 were slightly corrected by an amount of 50 K in $T_\mathrm{eff}$.  Fig. 5 to 7 show the distribution of the input 
parameters versus final parameters.
Some 44\% of the temperatures were corrected, 11\% of
the gravities and 7\% of the metallicities. This confirms that temperature is a very  
delicate parameter to estimate in 
spectroscopy and that its influence on the spectral profile is predominant. On the contrary,
the correction of metallicities was insignificant, showing that the weighted mean of
the determinations found in the literature was satisfactory. Fig. 7 in paper I shows how the 211 standards of 
the library are distributed in the plane
($T_\mathrm{eff}$, [Fe/H]) by intervals of gravity. The run of TGMET on the standards showed that the best solar analogue in this sample is HD 186427.
\\

The final parameters are listed in Table 1 which includes the following columns : identifiers BD/HD and
HIP, spectral type, equatorial coordinates (ICRS, epoch J1991.25), V magnitude, parallax and standard error in mas, ICRS proper motions and standard error in mas/yr, ELODIE radial velocity, final atmospheric parameters ($T_\mathrm{eff}$, $\log g$, [Fe/H]), level of reliability of the atmospheric parameters from 1 (poor) to 4 (high), bolometric (or visual) absolute magnitude,   
distance (if the relative error on parallaxe is lower than 30\%),  distances at plus and minus one sigma,  
 components 
of the spatial velocity (U,V,W) with respect to the Sun.


\begin{figure}
\resizebox{8cm}{!}{\includegraphics{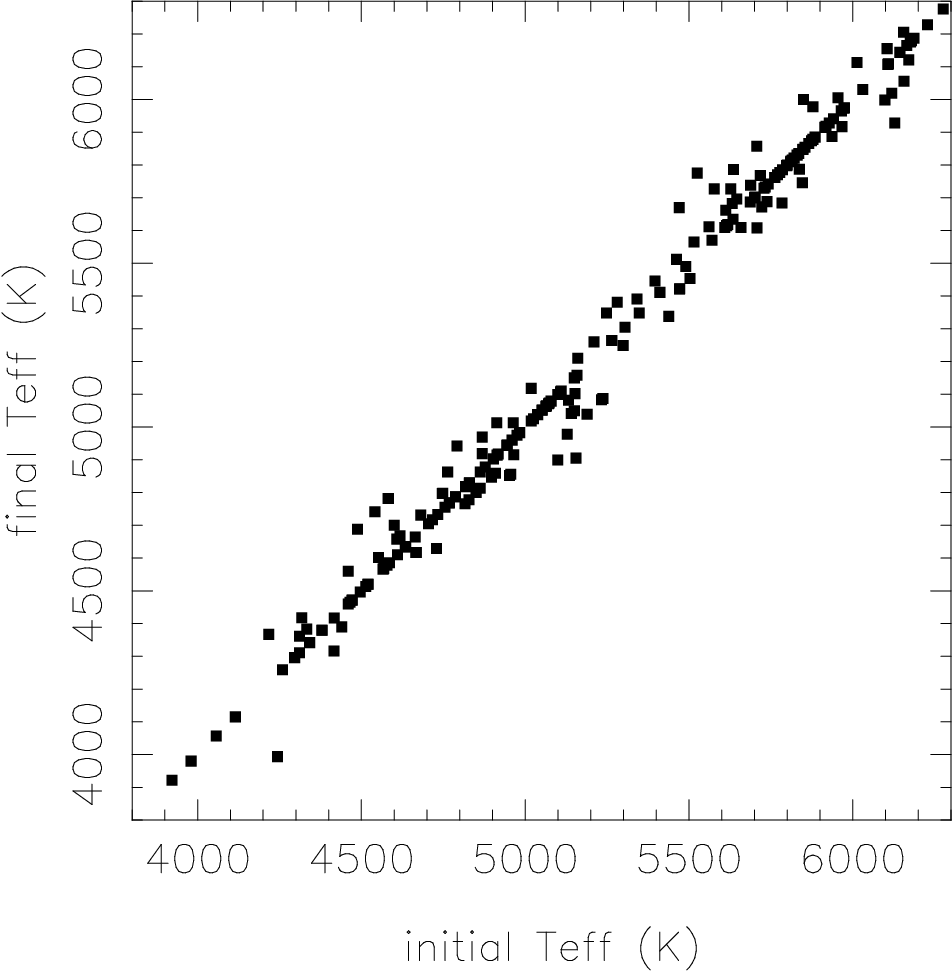}}
\caption{ Final $T_\mathrm{eff}$, obtained after 6 iterations of TGMET, versus input $T_\mathrm{eff}$, deduced
from the literature, for the 211 standards of the library.} 
\label{}
\end{figure}

\begin{figure}
\resizebox{8cm}{!}{\includegraphics{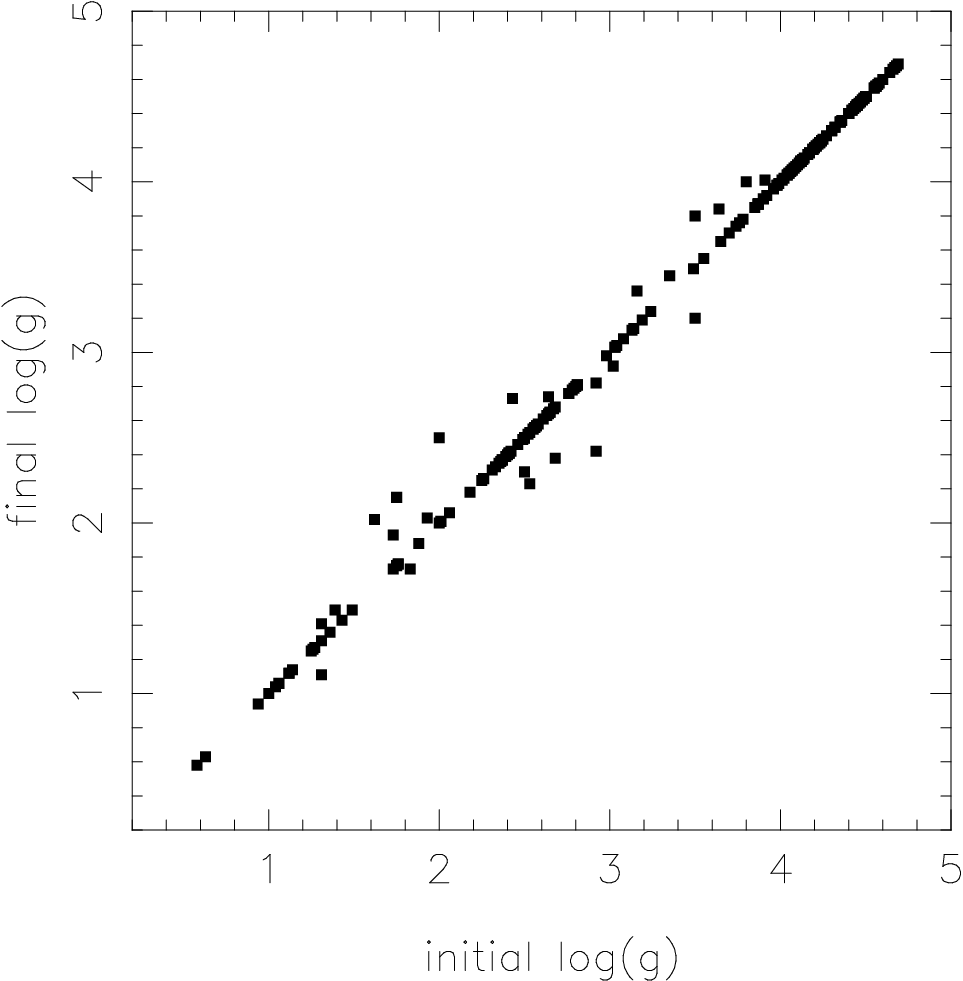}}
\caption{ Final $\log g$, obtained after 6 iterations of TGMET, versus input $\log g$, deduced
from the literature, for the 211 standards of the library.}
\label{}
\end{figure}

\begin{figure}[!h]
\resizebox{8cm}{!}{\includegraphics{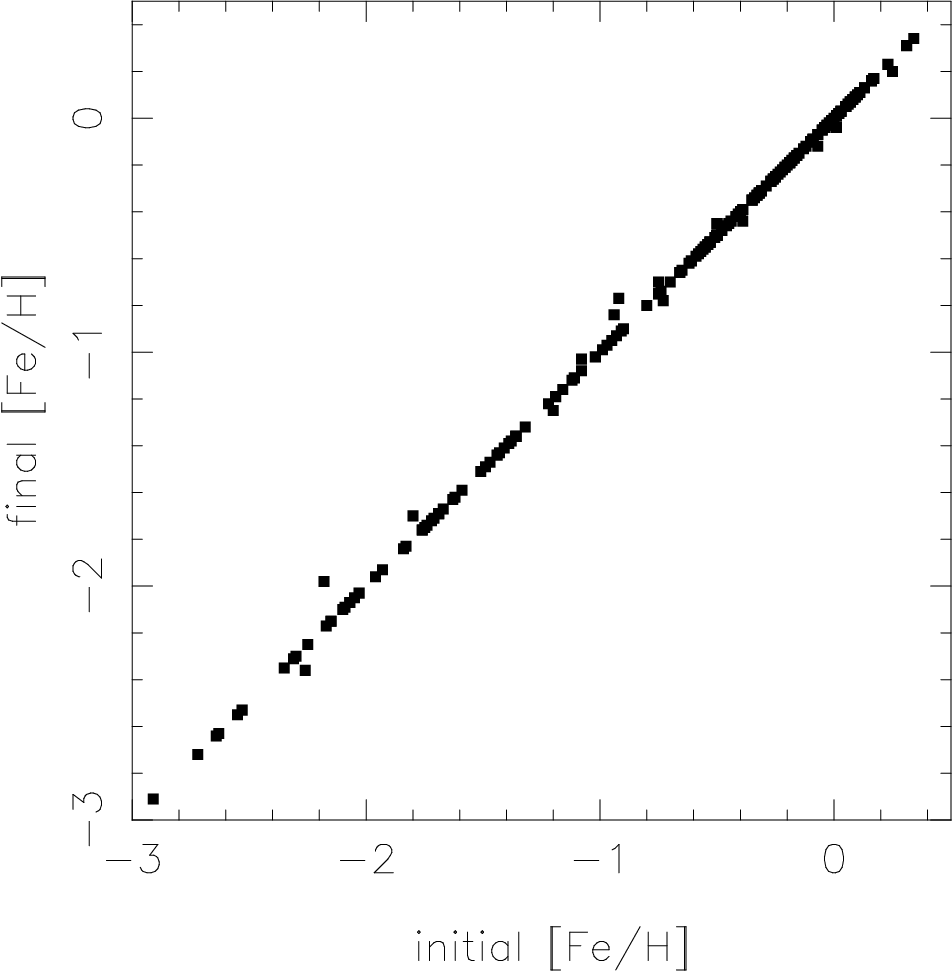}}
\caption{ Final [Fe/H], obtained after 6 iterations of TGMET, versus input [Fe/H], deduced
from the literature, for the 211 standards of the library.}
\label{}
\end{figure}

\begin{acknowledgements}
Our thanks go to all the observers who let us include in the library their spectra taken with ELODIE. We are grateful to Claude Catala and Jean-Claude Bourret for making some observations especially for this program.
\end{acknowledgements}

\end{document}